\pdfoutput=1
\documentclass[
    ,final            % use final for the camera ready runs
  ]
  {aipproc}

\usepackage{bibtex_abbrev}
\usepackage{natbib}
\usepackage{amssymb}

\layoutstyle{8x11double}

\begin{document}

\title{VERITAS Observations of the Coma Cluster of Galaxies}



\author{Jeremy S. Perkins}{ address={Fred Lawrence Whipple
    Observatory, Harvard-Smithsonian Center for Astrophysics, Amado,
    AZ 85645, USA},altaddress={jperkins@cfa.harvard.edu} }

\author{The VERITAS Collaboration}{ address={For a full list of
    authors see http://veritas.sao.arizona.edu } }

\begin{abstract}
  Clusters of galaxies are one of the few prominent classes of objects
  predicted to emit gamma rays not yet detected by satellites like EGRET or
  ground-based Imaging Atmospheric Cherenkov Telescopes (IACTs). The
  detection of Very High Energy (VHE, E > 100 GeV) gamma rays from
  galaxy clusters would provide insight into the morphology of
  non-thermal particles and fields in clusters. VERITAS, an array of
  four 12-meter diameter IACTs, is ideally situated to observe the
  massive Coma cluster, one of the best cluster candidates in the
  Northern Hemisphere. This contribution details the results of
  VERITAS observations of the Coma cluster of galaxies during the
  2007-2008 observing season.
\end{abstract}

\maketitle

%%%%%%%%%%%%%%%%%%%%%%%%%%%%%%%%%%%%%%%%%%%%
%% MAINMATTER
%%%%%%%%%%%%%%%%%%%%%%%%%%%%%%%%%%%%%%%%%%%%

\section{Introduction}

Galaxy clusters are the largest gravitationally bound objects in the
universe. A significant fraction of observable matter is embedded in
galaxy clusters which can have masses up to $10^{15}M_\odot$. Due to
their size and breadth of astrophysical phenomenon they have been
studied in almost every waveband. Some of the radio and X-ray
observations provide direct evidence for non-thermal particles and
magnetic fields within the Intra Cluster Medium (ICM), pointing to the
possibility of Very High Energy (VHE, E > 100 GeV) emission
\cite{Neumann:2003vl,Tribble:1993zl}.

\section{Non-Thermal Emission from Clusters of Galaxies}

Observational evidence in the X-ray, radio and extreme ultraviolet
indicates the presence of a population of non-thermal particles within
clusters. Synchrotron emission in the form of radio halos directly
point to highly energetic electrons.  These electrons are probably
accelerated to relativistic energies by shocks and mergers. In fact,
most clusters exhibit evidence for recent merger events. Individual
cluster members such as Active Galactic Nuclei and supernovae could
also accelerate electrons to these energies \cite{Berrington:2004sf}.
If these same processes are efficient at accelerating other particles,
there might also be a population of highly energetic protons. Protons
produced in this fashion will diffuse and remain within the cluster
for longer than a Hubble time due to their low energy loss rate and
the sheer size of the cluster.  Electrons however will cool much
faster and thus provide an instantaneous snapshot of the acceleration
processes going on within a cluster \cite{Volk:1996eu}. Both protons
and electrons can produce VHE emission. Electrons via Inverse Compton
off the Cosmic Microwave Background (CMB) and protons through hadronic
cascades. Starting with these acceleration mechanisms there are
multiple paths that generate non-thermal radio, X-ray and gamma-ray
emission, but protons only produce gamma rays through hadronic
cascades \cite{Pfrommer:2008}. Thus the observation of gamma-ray
emission produced via hadronic cascades can elucidate the
astrophysical processes present in galaxy clusters which is critical
in understanding cosmological structure formation. Furthermore, the
detection of VHE gamma rays from a galaxy cluster, combined with
multi-wavelength observations, would provide a detailed understanding
of the morphology of non-thermal particles and fields in the cluster.

There have been several predictions for gamma-ray emission from
clusters
\cite{Berrington:2004sf,Volk:1996eu,Keshet:2003nx,Gabici:2004hc}. Each
of these predictions are strongly dependent on the efficiency of the
emission processes. However, galaxy clusters are one of the few
objects predicted to emit gamma-ray radiation that have yet to be
detected by satellites like EGRET or ground-based IACTs
\cite{reimer.2003,Perkins:2006dq,Domainko:2007sf}. Only a small
fraction of the energy deposited into the ICM in a merger event need
be converted to non-thermal energy for there to be a detectable VHE
signal.  However, some authors have also stated that the probable
emission from galaxy clusters is well below the sensitivity of the
modern Imaging Atmospheric Cherenkov Telescopes (IACTs)
\cite{Kusnir.2008,Blasi:2007la}. Either the assumptions on the
efficiency of non-thermal energy conversion are overly optimistic or
there is something unknown about the non-thermal processes taking
place inside clusters of galaxies.

\section{The Coma Cluster}

The Coma cluster is a nearby cluster of galaxies which is well-studied
at all wavelengths
\cite{Neumann:2003vl,Tribble:1993zl,Biviano:1998pb}.  It is at a
distance of $\sim100$Mpc $(z\sim0.023)$ and has a mass of $2 \times
10^{15} M_\odot$.  The thermal X-ray plasma has a temperature of 8.25
keV.  In many ways, the Coma cluster is the best candidate for VHE
gamma-ray emission.  The radio halo, Coma C, is located within the
cluster, indicating the presence of non-thermal electrons.  In the
X-ray, there is excess hard emission seen by BeppoSAX
\cite{fusco-femiano.2004}, RXTE \cite{Rephaeli:2002bh} and INTEGRAL
\cite{Eckert:2007th} above the standard thermal component, although
the BeppoSAX detection is disputed. This emission could be interpreted
as inverse Compton scattering of CMB photons by electrons. Secondary
electrons might also contribute to a UV excess in the 130-180 eV range
\cite{bowyer.1998,Bowyer:2004dk}. There is evidence
\cite{Neumann:2003vl} of a recent merger event between the central
galaxies (NGC 4889 and NGC 4874) which might be responsible for the
radio emission.  Under this assumption, \citet{Berrington:2004sf}
argue that Coma should be detectable by modern IACTs if protons are
accelerated with the same efficiency as electrons. Previous VHE
gamma-ray observations of Coma by H.E.S.S. resulted in upper limits of
3.7\% of the Crab for point-like emission (above 1 TeV).  They also
reported an upper limit of 65.6\% of the Crab flux (above 2 TeV) for
extended emission centered on the core.  \cite{Domainko:2007sf}.

\section{Overview of VERITAS}
VERITAS is an array of four 12 m diameter IACTs located in Southern
Arizona ($31^\circ$ 40' 30'' N, $110^\circ$ 57' 08'' W) at an
elevation of 1268 m.  Fully operational since the Spring of 2007,
VERITAS is sensitive in the energy range from 100 GeV to greater than
30 TeV and has a 3.5 degree field of view.  VERITAS has an energy
resolution of $15\% - 20\%$ and a $0.1^\circ$ angular resolution (per
event, 68\% containment). Under normal operating conditions VERITAS
can detect a source at 5\% of the Crab Nebula's flux in $2.5$ hours
and 1\% in $50$ hours (depending on the spectrum). A detailed
description of the VERITAS instrument and technique is available in
\cite{VERITAS-Collaboration:-E.Hays:2007fe}, \cite{Holder:2006pi} and
\cite{Holder:2008}.

\section{Data Set}

Observations of the Coma cluster were made with the full telescope
array from March through May, 2008.  After quality selection, the
total exposure is 18.6 hours live time. FIR measurements indicate that
the weather was ideal over the entire observing period.  All
observations were performed in a small range of zenith angles
($9-20^\circ$).  Data were taken on moonless nights in ``wobble''
mode, where the telescopes are offset from the source by $0.5^\circ$,
to allow for simultaneous background estimation using events in the
same field of view.

\section{Results and Outlook}

\begin{figure}
  \includegraphics[height=.3\textheight]{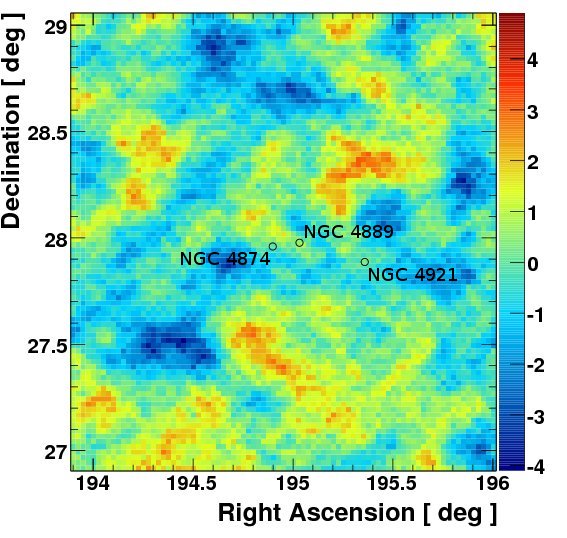}
  \caption{Significance Map of the Coma cluster in VHE Gamma Rays.  No
    evidence for point like or extended emission is seen.  The
    locations of three major galaxies located in the cluster are
    included.}
  \label{fig:sigma_map}
\end{figure}

\begin{figure}
  \includegraphics[height=.3\textheight]{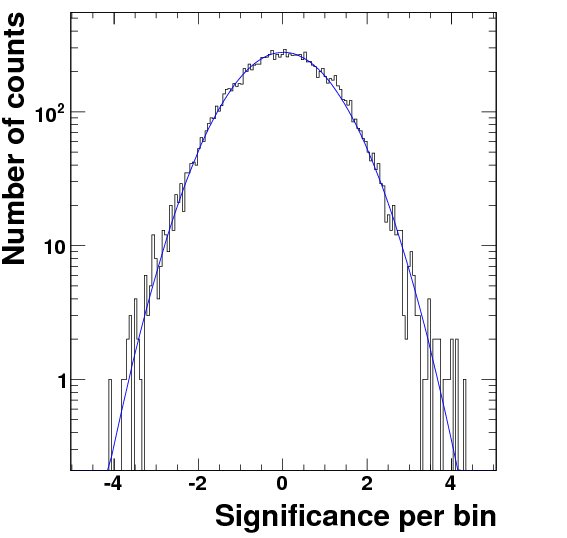}
  \caption{Significance distribution over the field of view shown in
    Figure \ref{fig:sigma_map}.  This distribution was fit with a
    gaussian using the width and mean as free parameters. This fit
    resulted in a width of 1.0 and mean of 0.0.}
  \label{fig:sigma_dist}
\end{figure}

\begin{table}
\begin{tabular}{lrrrr}
\hline
\tablehead{1}{r}{b}{Source}
& \tablehead{1}{r}{b}{Location}
& \tablehead{1}{r}{b}{ON/OFF/Excess}
& \tablehead{1}{r}{b}{Upper Limit\\(ph m$^{-2}$ s$^{-1}$)}
& \tablehead{1}{r}{b}{Upper Limit\\(Crab Flux)}\\
\hline
Center (Point)    & 12 59 48.7 +27 58 50 & 204/2340/-8.7  & $8.82\times10^{-9}$ & $0.7\%$  \\
Center (Extended) & 12 59 48.7 +27 58 50 & 1419/2865/13.5 & $\sim2\times10^{-8}$ & $\sim3\%$ \\
NGC 4889          & 13 00 08.03 +27 58 35.1 & 218/2310/7.0   & $1.01\times10^{-8}$ & $0.8\%$  \\
NGC 4874          & 12 59 35.91 +27 57 30.8 & 197/2419/-18.5 & $7.76\times10^{-9}$ & $0.6\%$  \\
NGC 4921          & 13 01 26.23 +27 53 08.5 & 184/2315/-7.0  & $8.47\times10^{-9}$ & $0.7\%$  \\
\hline
\end{tabular}
\caption{Upper Limits}
\label{tab:results}
\end{table}

The data are processed using the standard cleaning and analysis
methods. For details on these, see \cite{Daniel:2007kx}. Event
selection is performed using weak source cuts optimized for a 0.3\%
Crab flux source. Upper limits (99\% confidence level) on the integral
flux above 300 GeV are calculated using the method of Helene
\cite{helene.1983}.  For a point-like core region (r < $0.115^\circ$),
there were 204 ON and 2340 OFF events resulting in an excess of
-8.7. The upper limit on the gamma-ray flux from the core region is
$8.82\times10^{-9}$ ph m$^{-2}$ s$^{-1}$ ($0.7\%$ of the Crab
flux). The preliminary upper limit on a mildly extended region (r <
$0.300^\circ$) is $\sim2\times10^{-8}$ ph m$^{-2}$ s$^{-1}$ ($\sim3\%$
of the Crab flux).  There are 1419 ON and 2865 OFF events resulting in
a excess of 13.5 for this extended region.  Upper limits on selected
cluster members (NGC 4889, NGC 4874 and NGC 4921) are summarized in
Table \ref{tab:results}.  In addition to these results for specific
locations, a search in the field of view for any signals is also
performed (see Figure \ref{fig:sigma_map}).  No significant sources of
gamma-rays are seen and the distribution of significances is well fit
by a Gaussian centered on 0.0 with a width of 1.0 (see Figure
\ref{fig:sigma_dist}).

In summary, VERITAS observed the Coma galaxy cluster for approximately
19 hours in Spring 2008.  No evidence for point-source emission was
observed within the field of view and a preliminary upper limit of
$\sim3\%$ of the Crab flux is given for a moderately extended region
centered on the core. Even though the search for TeV emission from
clusters of galaxies has not resulted in any detections, the outlook
for detecting gamma-ray emission from clusters is promising in light
of the recent launch of the Fermi satellite.

%%%%%%%%%%%%%%%%%%%%%%%%%%%%%%%%%%%%%%%%%%%%%%%%
%% BACKMATTER
%%%%%%%%%%%%%%%%%%%%%%%%%%%%%%%%%%%%%%%%%%%%%%%%

\begin{theacknowledgments}
  This research is supported by grants from the U.S. Department of
  Energy, the U.S. National Science Foundation, and the Smithsonian
  Institution, by NSERC in Canada, by PPARC in the UK and by Science
  Foundation Ireland.
\end{theacknowledgments}

\bibliographystyle{aipproc-custom}   % if natbib is available

\bibliography{/Users/jperkins/Documents/bibliography}

\end{document}